\documentclass[letterpaper,english,reprint, aps]{revtex4-1}
\usepackage[T1]{fontenc}
\usepackage[latin9]{inputenc}
\usepackage{xcolor}
\usepackage{verbatim}
\usepackage{float}
\usepackage{calc}
\usepackage{amsmath}
\usepackage{amssymb}
\usepackage{graphicx}



\usepackage{babel}
\begin{document}

\preprint{APS/123-QED}

\title{Magnetic field control of antiferromagnetic domain walls in a thermal gradient\\
}
\author{ R. Yanes$^1$,  M. Rodriguez Rosa$^1$ and L. Lopez-Diaz$^1$}
\email{ryanes@usal.es}

\affiliation{$^1$Department of Applied Physics, University of Salamanca, 37008 Salamanca,  Spain}

\begin{abstract}
An antiferromagnetic domain wall in a thermal gradient is found to experience a force towards colder regions upon the application of a uniform magnetic field along the easy axis. This force increases with the strength of the applied field and, for sufficiently high values, it overcomes the entropic force the that pushes wall towards the hotter regions. The force is proportional to the thermal gradient and it shows a linear dependence with the net magnetic moment of the domain wall induced by the field. The origin of this force lies on the increase of the domain wall reflectivity due the field-induced sizable break of antiferromagnetic order inside it, which turns it into an efficient barrier for magnons, which transfer linear momentum to the domain wall when they are reflected on it.\end{abstract}
\maketitle

\section{\label{sec:intro}introduction:}
Although they have played mostly a passive role in devices so far, antiferromagnetic materials (AF) present some appealing properties that, together with the development of different techniques to manipulate and detect the orientation of their N\'eel vector, confer them a major role in the future of spintronics \cite{baltz_antiferromagnetic_2018}.

Theoretical investigations have shown that domain walls (DW) in AF can move much faster than their ferromagnetic counterparts due to their low effective mass and the absence of a Walker breakdown limit \cite{gomonay_high_2016, selzer_inertia-free_2016}. Like in ferromagnets, current-driven DW motion in AF can be achieved via spin-transfer torque \cite{swaving_current-induced_2011,yamane_combined_2017} and also by spin-orbit torque when the current flows through an adjacent heavy metal layer \cite{gomonay_high_2016,shiino_antiferromagnetic_2016}. Both methods are currently being extensively investigated, but other approaches that do not require charge currents are also worth exploring, such as using field gradients \cite{tveten_intrinsic_2016,yuan_classification_2018}, oscillating fields \cite{li_rotating_2020}, asymmetric field pulses \cite{gomonay_manipulating_2016}, anisotropy gradients \cite{wen_ultralow-loss_2020}, magnonic currents \cite{tveten_antiferromagnetic_2014,kim_propulsion_2014} or thermal gradients \cite{selzer_inertia-free_2016, wu_antiferromagnetic_2016}. 

Regarding this last method, it was demonstrated theoretically that an AF DW in a thermal gradient moves towards hotter regions, as it is the case in ferromagnets. This motion was explained using thermodynamic arguments, namely that it maximizes the entropy of the system or, equivalently, it minimizes its free energy \cite{selzer_inertia-free_2016}. An analytical equation for the domain wall motion was derived based on this so-called entropic torque \cite{selzer_inertia-free_2016, cheng_antiferromagnet-based_2018}, following a similar procedure than for ferromagnets \cite{schlickeiser_role_2014,kim_landau-lifshitz_2015}. It was also suggested that the Brownian motion of a DW or any other soliton in an AF leads to a small small drift towards colder regions, a phenomenon called thermophoresis \cite{kim_thermophoresis_2015}. Another possible mechanism for thermally driven DW motion is the transfer of angular momentum from magnons to the DW when passing through it \cite{yan_all-magnonic_2011,kim_landau-lifshitz_2015}, leading to an adiabatic torque on the DW wall that drives it against the magnon flow, i.e., towards hotter regions \cite{hinzke_domain_2011}. In compensated AF, thermal magnons do not carry angular momentum on average \cite{ritzmann_thermally_2017} and, therefore, this mechanism has, in principle, no effect on the DW. However, in ferrimagnets, it has been shown recently \cite{donges_unveiling_2020} that, above the Walker breakdown, the magnonic adiabatic torque changes sign around the compensation temperature, driving the DW towards the hot end above it and towards the cold end below it.
On the other hand, it has been proposed recently \cite{shen_driving_2020} that, in uniaxial AF, applying a field along the easy axis enhances magnon reflection, which contributes to the
thermally driven domain-wall motion . 

%\bl{There are also experimental evidences of the existence of thermally induced spin currents in antiferromagnetic materials like in the conducting magnetite thin films\cite{ramos_observation_2013}, multilayers of [Fe$_2$O$_3$/Pt]$_n$\cite{ramos_unconventional_2015}, or  in  insulating heterostructures as Cr$_2$O$_3$/Pt\cite{seki_thermal_2015}, MnF$_2$/Pt \cite{wu_antiferromagnetic_2016}, bilayers of [NiO/M] being M=Pt, Ta, IrMn or Py (permalloy)\cite{holanda_spin_2017}, Si/Py/NiO/Pt\cite{ribeiro_spin_2019} and epitaxially growth  FeF$_2$(110) thin film\cite{li_spin_2019}. Some of those studies were performed varying the magnitude of the applied magnetic field \cite{wu_antiferromagnetic_2016, seki_thermal_2015,holanda_spin_2017,li_spin_2019} showing an increment on the spin Seebeck signal with the applied field. Those experimental results, open the door to the question if an applied field will tune the DW speed in an  thermally induced DW motion in antiferromagnetic materials. }

From the experimental side, there is already strong evidence of thermally excited spin currents in different AF materials, such as  Cr$_2$O$_3$\cite{seki_thermal_2015}, MnF$_2$ \cite{wu_antiferromagnetic_2016}, NiO \cite{holanda_spin_2017,ribeiro_spin_2019}  or FeF$_2$\cite{li_spin_2019}, tipically detected by Inverse Spin Hall Effect on an adjacent heavy metal layer. Some of these studies were performed under an external magnetic field \cite{ seki_thermal_2015,holanda_spin_2017,li_spin_2019} showing an increment of the spin Seebeck signal with the applied field amplitude. These experimental results, together with the theoretical works mentioned before, lead to the question of whether an applied field influences thermally induced DW motion in AF materials.

In this work we use atomistic spin simulations to investigate thermally driven DW motion in a biaxial AF under an external magnetic field along the easy axis. We find that, as the field is increased, the DW gradually shifts from moving towards the hot end to moving towards the cold one, changing sign roughly mid-way to the spin-flop transition. Furthermore, the DW velocity dependece on the applied field collapses to a single curve when normalized to the temperature gradient. By performing a systematic frequency analysis with monocromatic excitations we correlate the DW displacement with magnon reflection. The study reveals that the effect of the applied field is twofold. On one hand, it increases the gap between the two magnon branches, allowing for the propagation of magnons of lower frequency that are more effectively reflected by the DW. On the other, the DW reflectivity itself increases, in good agreement with \cite{shen_driving_2020}, due to the sizable net magnetic moment of the DW, which significantly breaks antiferromagnetic order, thus becoming a barrier for magnons. 

The manuscript is organized as follows. In Section \ref{sec:model} the atomistic model used for our investigation is explained and some relevant computational details are given, whereas the main results are presented and discussed in Section \ref{sec:results}. Firstly, the influence of the applied field on DW motion in a thermal gradient is characterized and, later on, a systematic analysis with monochromatic excitations is presented for a better understanding of the results. The main conclusions of our work are presented in Section \ref{sec:conclusions}.

\section{\label{sec:model}Atomistic model}
We use an atomistic spin model for our bipartite AF based on the following Hamiltonian:

\begin{equation}
\label{eq_hamiltonian}
\mathcal{H}=- \sum_{<i,j>}J\,\vec{s}_{i}\cdot\vec{s}_{j}-\sum_{i}\left( K_{x}\,s_{i,x}^2
+K_{y}\,s_{i,y}^2+\mu_{\rm s}\,\vec{s}_{i}\cdot\vec{B}_{ext} \right)
\end{equation}

\noindent where $\vec{s}_i$ is the normalized magnetic moment ($\vec{s}_i=\vec{\mu}_i/\mu_{\rm s}$), $J$ is the exchange constant, $K_x$ are $K_y$ are the primary and secondary anisotropy constants ($K_x > K_y$), respectively, and $B_{ext}$ is the applied field. The dynamics of the system is given by the stochastic Landau-Lifshitz-Gilbert equation

\begin{equation}
\label{eq_SLLG}
\frac{\partial\vec{s}_{i}}{\partial t}=-\frac{\gamma}{(1+\alpha^{2})\mu_{{\rm s}}}\vec{s}_{i}\times\left(\vec{H}_{i}+\alpha\,\vec{s}_{i}\times\vec{H}_{i}\right),
\end{equation}

\noindent where $\gamma$ is the gyromagnetic ratio, $\alpha$ is the damping parameter and the $\vec{H}_{i}=-\frac{\partial\mathcal{H}}{\partial\vec{s}_{i}}+\vec{\zeta}_{i}(t)$. We take into account thermal fluctuations at finite $T$ by adding the stochastic term $\vec{\zeta}_{i}(t)$ obeying the statistical properties of white noise \cite{brown_thermal_1963,garcia-palacios_langevin-dynamics_1998}. 

The system under study is schematically represented in Fig.\ref{fig:DWmot}(a). We consider a bipartite AF in the shape of a nanowire of dimensions $L_x \times L_y \times L_z = 512\,a\times 16\,a \times 16\,a$, where $a$ is the lattice constant. A DW separating two antiparallel domains magnetized along the easy axis ($x$) is initially placed at the center of the nanowire and relaxed at $T=0$. Once equilibrium is reached, the dynamics of the system under the influence of both a uniform thermal gradient with its hot end on the left and a magnetic field along the easy axis is computed by integrating (eq. \ref{eq_SLLG}) numerically using Heun's scheme \cite{suli_introduction_2003} with a fixed time step $\Delta t = 2.0\times 10^{-3}\,\tau$, where $\tau=\mu_{s}/(\gamma\cdot J)=56.35$ fs. The following material parameter values were considered: $a=0.5$ nm, $J=-2\times 10^{-21}$ J, $\mu_{\rm s}=2.14\,\mu_B$, $K_x=2\times 10^{-22}$ J, $K_y=0.1\,K_x$, $\alpha=1\times 10^{-2}$. 
%%------------------------------ previos-----------

\section{\label{sec:results}Results and discussion}

\begin{figure}[ht]
	\begin{centering}
		(a)\includegraphics[width=0.45\textwidth]{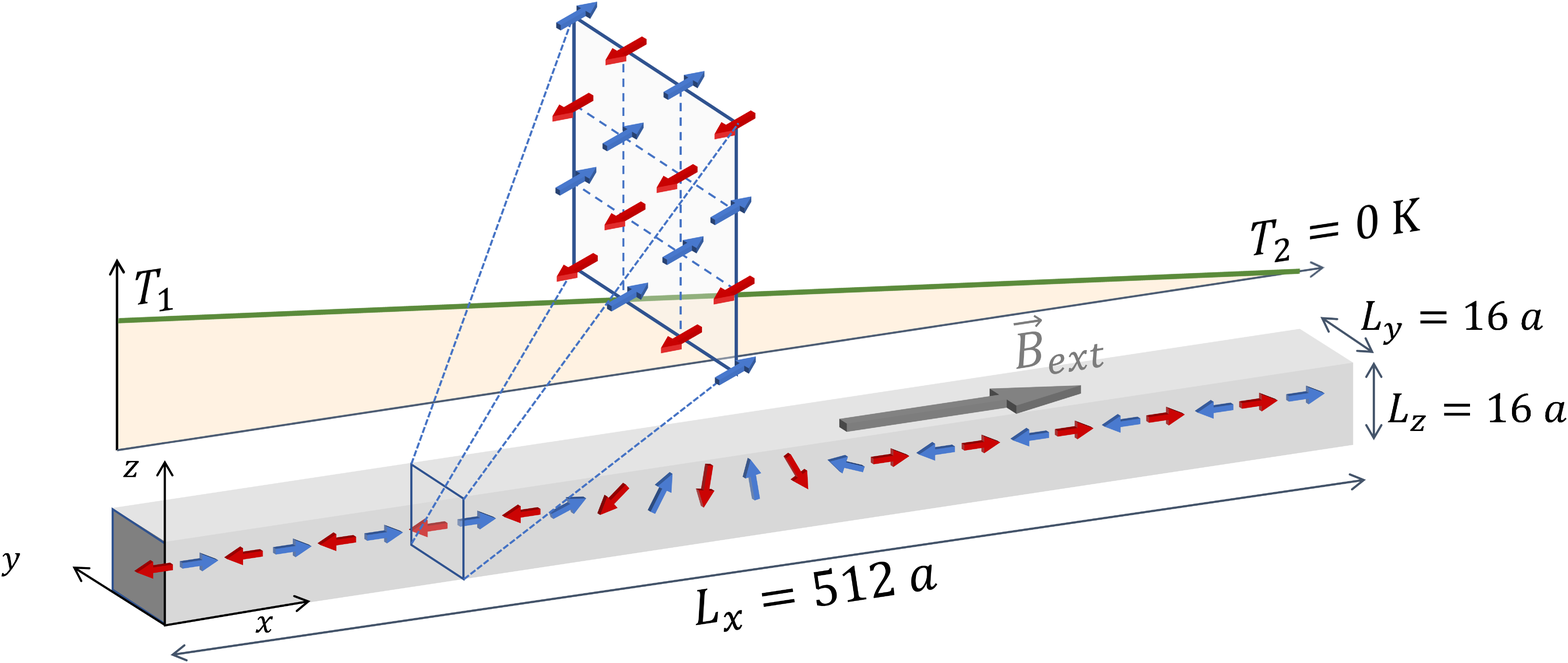}
		(b)\includegraphics[width=0.45\textwidth]{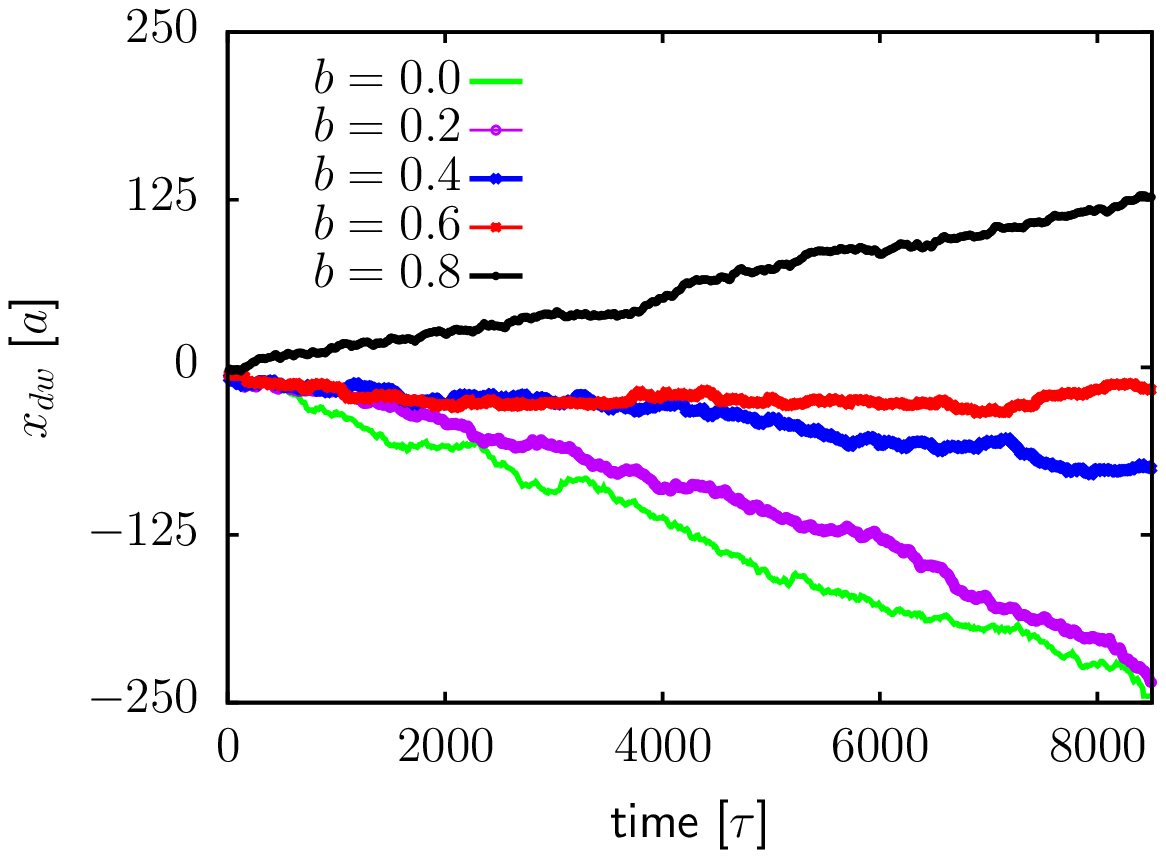}
		(c)\includegraphics[width=0.45\textwidth]{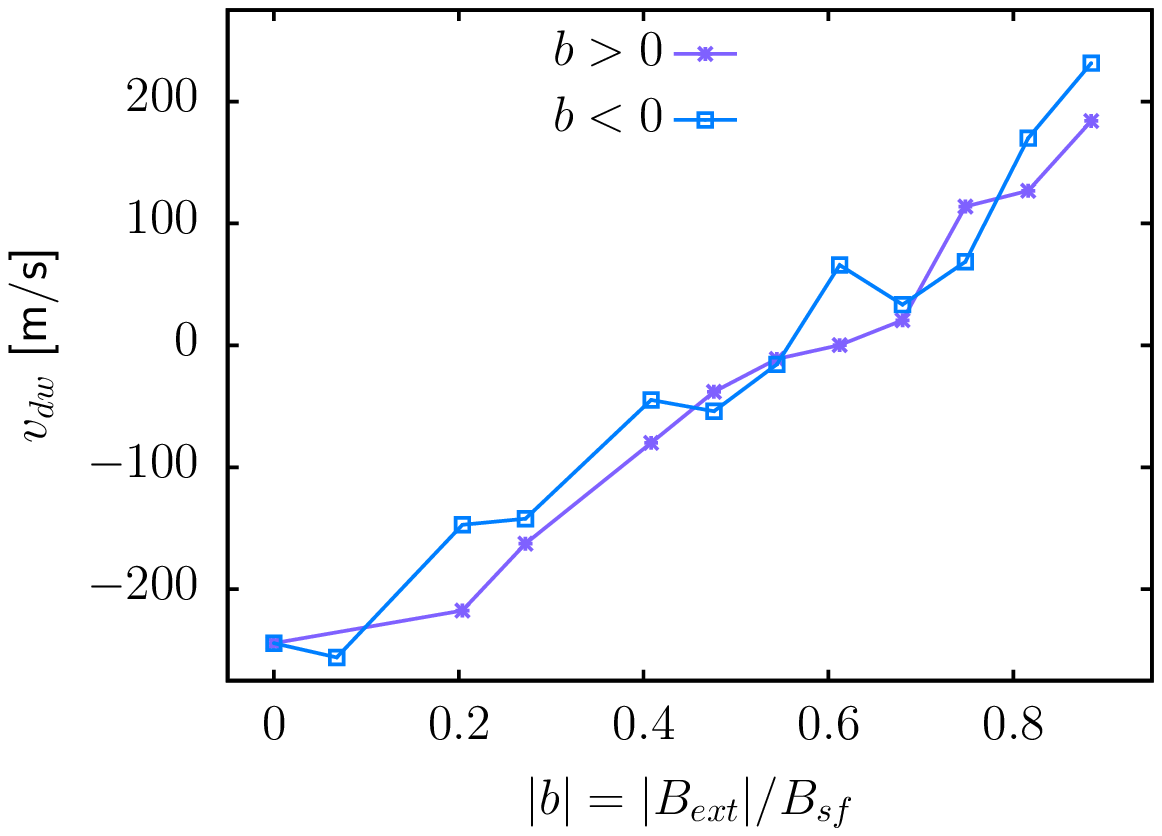}
		\par\end{centering}
	\centering{}\caption{\label{fig:DWmot} (a) Sketch of the system under study. The red and blue arrows represent the spins of magnetic sublattice 1 and 2, respectively. For a thermal gradient of $\frac{dT}{dx} = 0.194$ K/nm, (b) time evolution of the DW position ($x_{dw}$) for different values of the applied field $b = B_{ext}/B_{sf}$ and (c) average DW velocity ($v_{dw}$) as a function of the applied field strength $|b|$ for both positive (purple) and negative (blue) values.}
\end{figure}

Fig.\ref{fig:DWmot}(b) shows the time evolution of the DW position for different values of the applied field, which is normalized to the spin-flop transition value with $K_y=0$, $b=B_{ext}/B_{sf}$, where  $B_{sf}=\sqrt{2B_{ex}B_{an}-B_{an}^2}$, being $B_{ex}=\frac{6J}{\mu_{s}}$ and $B_{an}=\frac{2K_x}{\mu_{s}}$. 

Without the external field ($b=0$, green line in Fig. \ref{fig:DWmot}(b)) the DW moves leftwards, towards the hotter region, as a result of the force exerted on the DW due to the entropic torque \cite{selzer_inertia-free_2016}, but the behaviour is substantially modified in the presence of magnetic field. We observe a gradual shift from the negative slope at $b=0$ towards positive slope for large fields, which clearly indicates that the magnetic field favours DW motion towards colder regions. This is confirmed if we now compute the DW velocity, averaged over a time window of $t_{sim}=8.4\times 10^3\, \tau$ and over 5 realizations, as a function of the applied field. The results are presented in Fig. \ref{fig:DWmot}(c) for both positive (purple) and negative (blue) fields. They prove that it is possible to reverse the motion of a AF DW inside a thermal gradient with a magnetic field and, furthermore, that its speed can be tuned with the strength of the field. This is the main result of our work.

%%Figure \ref{fig:DWmot} (a) Sketch of the system under study. The red and blue arrows represent the spins of magnetic sublattice 1 and 2, respectively. For a thermal gradient of $\frac{dT}{dx} = 0.194$ K/nm, (b) time evolution of the DW position ($x_{DW}$) for different values of the applied field $b = B_{ext}/B_{sf}$ and (c) average DW velocity ($<v_{DW}>$) as a function of the applied field strength $|b|$ for both positive (purple) and negative (blue) values. 

%---------------------------figure 1---------------

We repeated our study for different values of the temperature gradient in order to check the soundness of our finding. The results are shown in Fig. 2(a), were the average DW velocity divided by the thermal gradient is plotted in the $y$ axis. As can be observed, all the tested cases tend to fall within in a single curve, which indicates that the effect of the applied field on the DW velocity is proportional to the temperature gradient, as it happens to be the case of the entropic torque \cite{selzer_inertia-free_2016}. The data in Fig. 2(a) are fitted to the following law

\begin{equation}
v_{dw}=\Big(A-C|b|^{n}\Big)\frac{dT}{dx}\label{eq:vdw}
\end{equation} 

\noindent giving $A=(1.37 \pm 0.11)\times 10^{-6} \, \rm{m}^2 \, \rm{s}^{-1} \,\rm{K}^{-1}$, $C=(3.05\times \pm 0.16)\times 10^{-6} \, \rm{m}^2 \,\rm{s}^{-1} \,\rm{K}^{-1}$ and $n=1.25\pm 0.15$. The coefficient $A$ represents the contribution of the entropic torque in absence of the applied field. According to \cite{selzer_inertia-free_2016}, $A=\frac{2\gamma\,a^3}{\mu_s\alpha}\frac{\partial A_{ex}}{\partial T}$ which, for our parameter values and assuming a linear temperature dependence of the micromagnetic exchange constant $A_{ex}$, gives $A=1.53\times 10^{-6}\, \rm{m}^2 \, \rm{s}^{-1} \,\rm{K}^{-1}$, in good agreement with the fitted value obtained from our data. 

%---------------------- Figure2-----------------------------
\begin{figure}
	\begin{centering}
		(a)\includegraphics[width=0.45\textwidth]{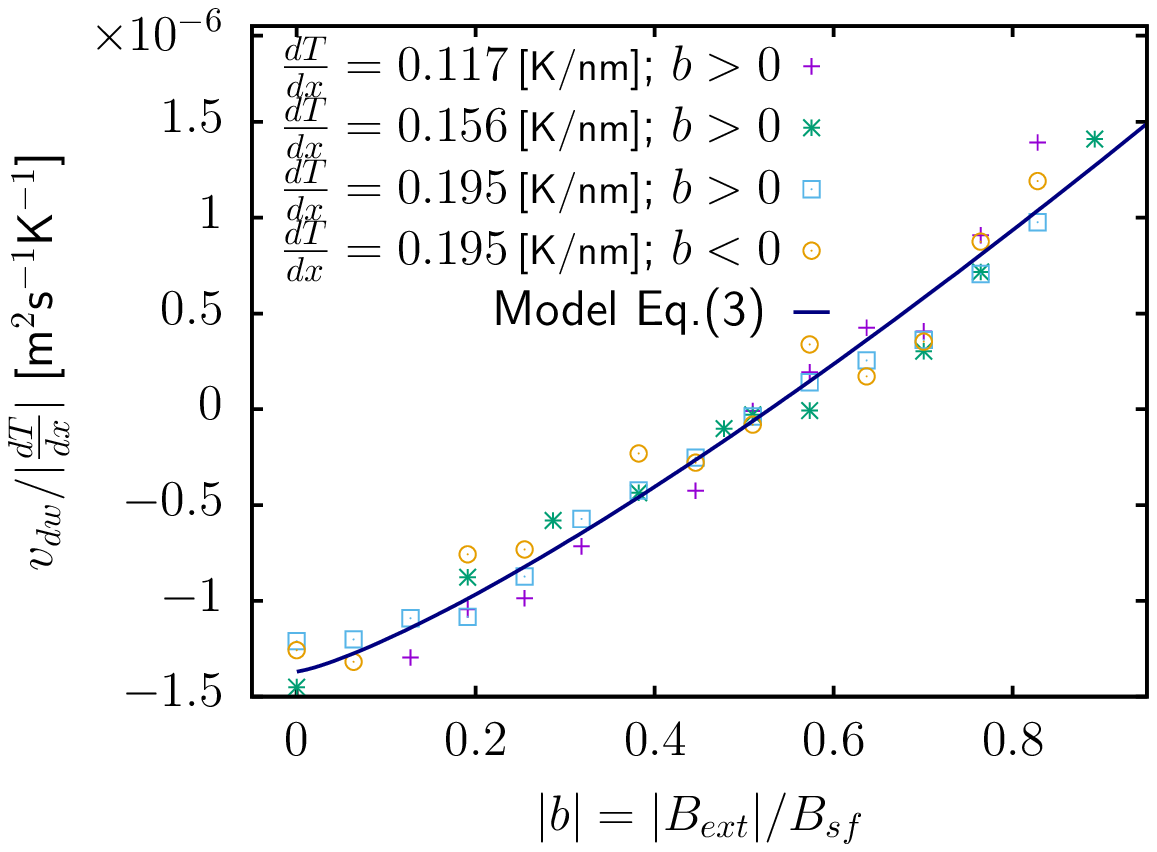}
		(b)\includegraphics[width=0.45\textwidth]{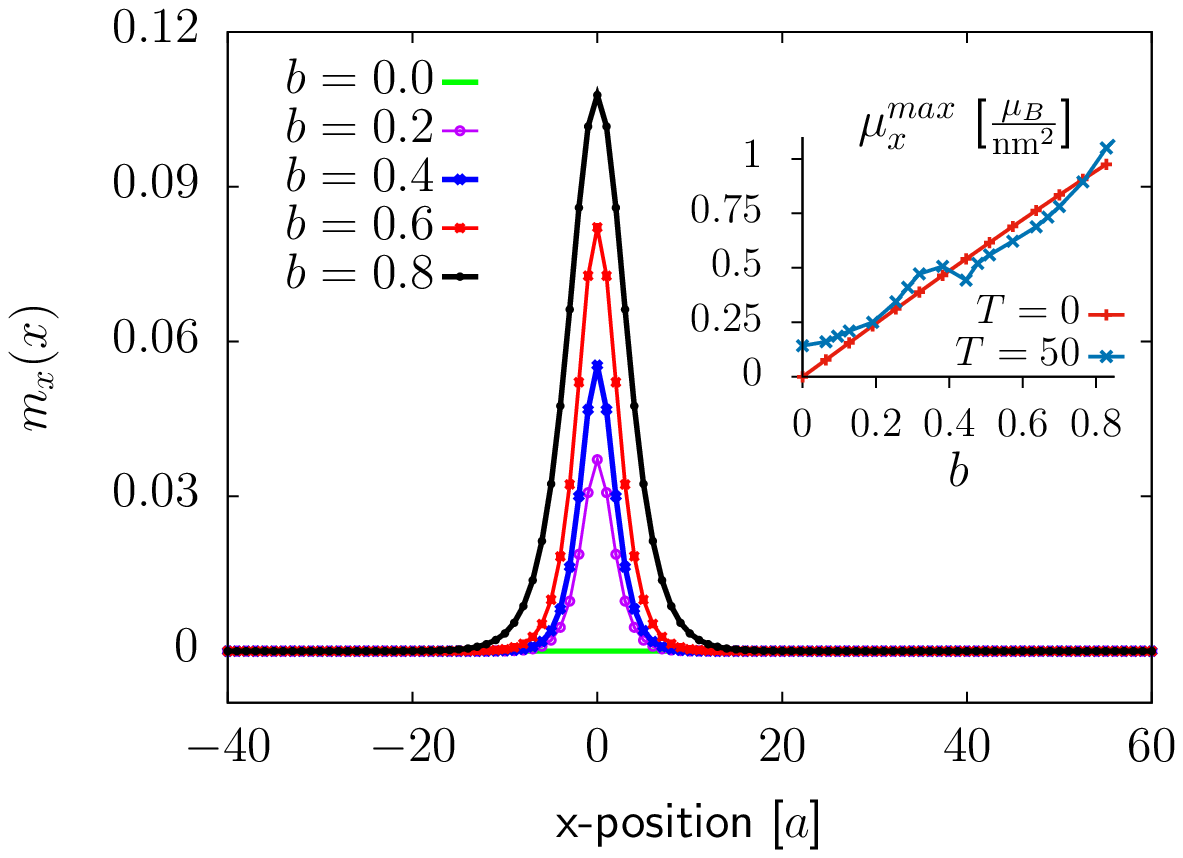}
		(c)\includegraphics[width=0.45\textwidth]{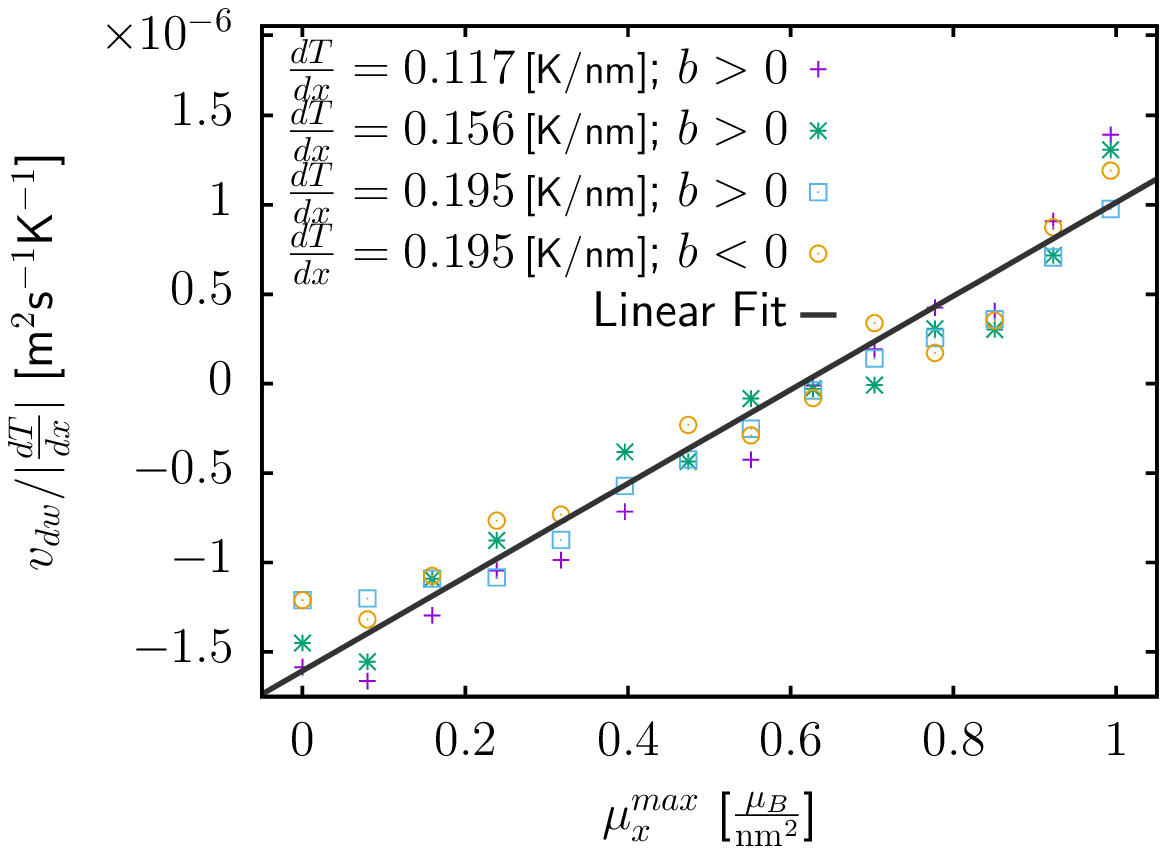}
	\end{centering}
	\caption{\label{fig:Vel-DWmx} (a) Average DW velocity divided by the strength of the thermal gradient as a function of the applied field. The line corresponds to the fit of the data to (\ref{eq:vdw}). (b) Profile of the net magnetization $m_x$ along the DW (centered at $x=0$) for different values of the magnetic field. Inset: DW maximum magnetic moment per unit surface as a function of the applied field. (c) Average DW velocity divided by the strength of the thermal gradient as a function DW maximum net magnetic moment per surface. The line corresponds to a linear fit of the data. }
\end{figure}
%--------------------------------------------------

It is natural to think that the effect of the magnetic field on DW motion is related to the change in its internal structure. Fig. 2(b) shows the profile of the net magnetic moment $m_x=\sum_i \vec{s}_{i,x}$ averaged over the cross section of the nanowire for a DW centered at $x=0$ in equilibrium at $T=0$  for different field values. As can be observed, whereas the AF domains are insensitive to the field ($m_x\approx 0$), a sizable net magnetization parallel to it inside the DW. Both the width and heigth of the $m_x$ distribution increase with the magnetic field. This is due to the fact that the spins inside the DW are oriented perpendicular to the field and, therefore, their susceptibility is much higher than for those inside the domains \cite{tveten_intrinsic_2016,sass_magnetic_2020}. In the inset of Fig. \ref{fig:Vel-DWmx}(b) we plot the net magnetic moment per unit area ($\mu_x^{max}=m_x^{max}\,\mu_B / a^2$) at the center of the DW, where it reaches its maximum value, as a function of the magnetic field, displaying a very similar dependence than the DW velocity (Fig. \ref{fig:Vel-DWmx}(a)). In fact, if we now plot the DW velocity as a function of $\mu_x^{max}$ we obtain an apparent linear dependence (Fig. \ref{fig:Vel-DWmx}(c)), which points out the central role played by the DW net magnetic moment in the the observed behaviour.

In the end, this sizable net magnetization means that a substantial deviation from AF ordering takes place inside the DW, which most likely will affect the way the DW interacts with the thermally excited magnonic current. Furthermore, the effect of the magnetic field on DW velocity is drastically reduced if we increase the damping parameter $\alpha$, which points to the magnonic nature of this effect. On the other hand, the magnetic field not only opens the gap between the two AF modes in the system \cite{rezende_introduction_2019}  but it has also been shown that it can modify to a large extent the propagation length of thermally excited magnons \cite{bender_enhanced_2017,lebrun_tunable_2018}. It is difficult to say, a priori, how each one of these aspects affect DW dynamics and, in particular, which one is responsible for the field-induced motion of the DW towards the cold reservoir. To elucidate this issue in what follows we investigate in detail how the DW interacts with monochromatic spin waves. 

%---------------------- Figure 3------------------------------
\begin{figure}[ht]
	\begin{centering}
		(a)\includegraphics[width=0.45\textwidth]{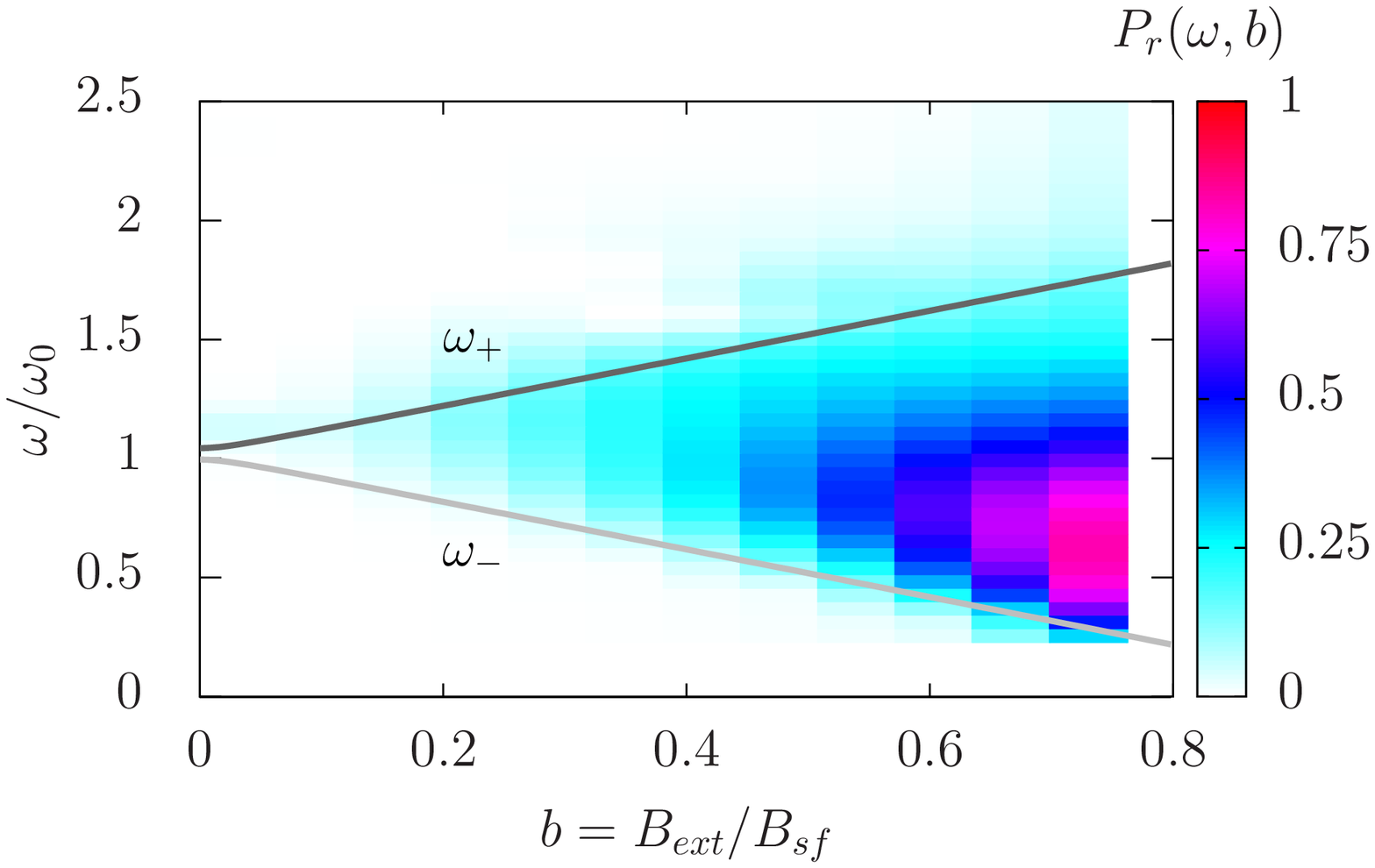}
		(b)\includegraphics[width=0.45\textwidth]{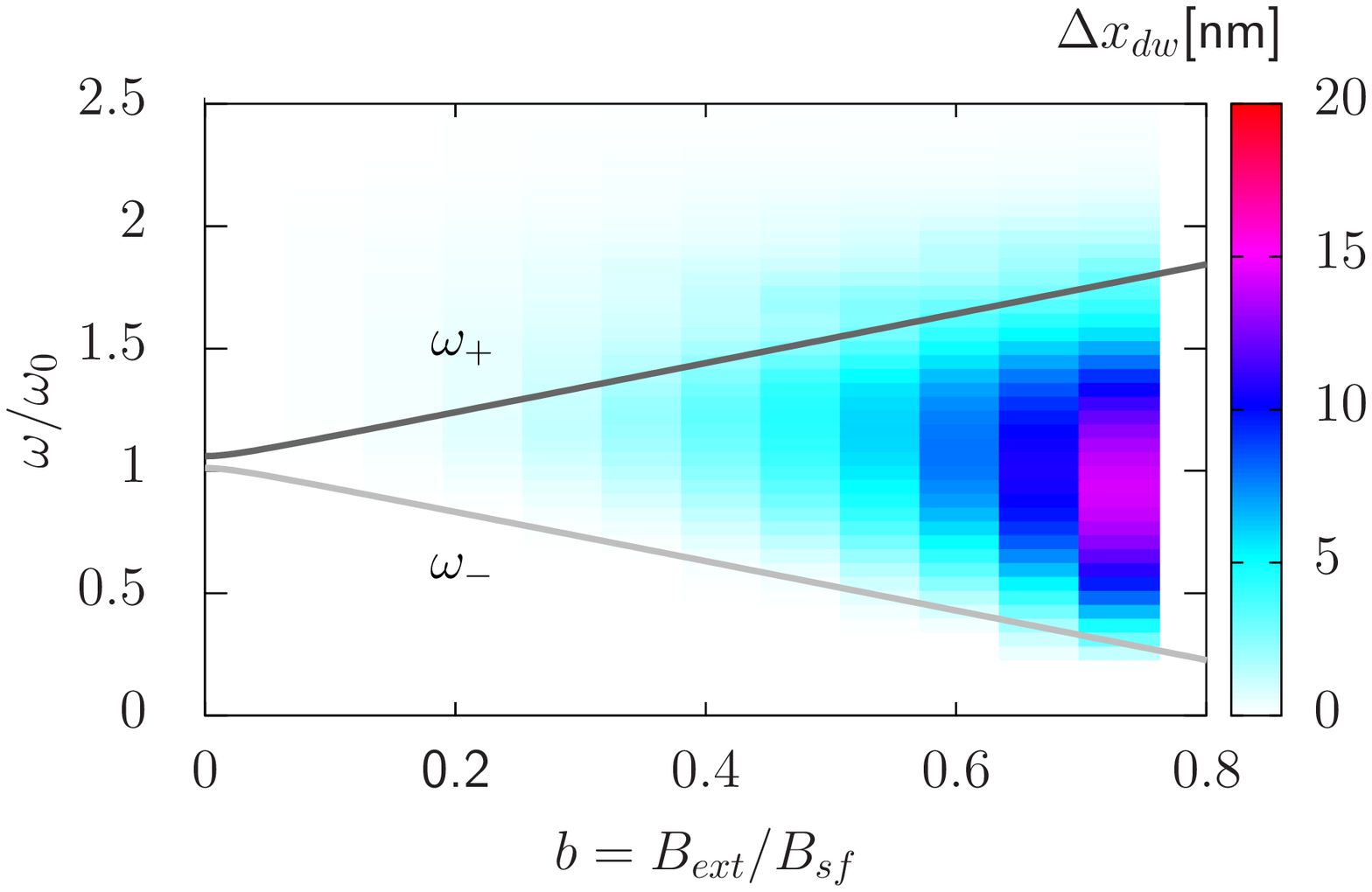}
	\end{centering}
	\caption{\label{fig:DWdis-Ref}(a) Reflected power $P_r$, and (b) DW displacement $\Delta x_{dw}$ as a function of applied field and frequency for a monochromatic AF excitation. Both applied field and frequency are normalized to $B_{sf}$ and $\omega_0=2\pi\gamma B_{sf}$, respectively.} 
\end{figure}
%%-----------------------------------------------------------

Monochromatic spin waves (SW) are injected in our system by imposing a sinusoidal excitation on the spins at the left edge ($s_x=\sqrt{1-A^2},\,s_y=A\sin(\omega t), \,s_z=A\cos(\omega t)$, where $A=0.2$ and $\omega $ are the amplitude and frequency of the excitation, respectively. As it propagates along $x$ the SW interacts with the DW, which is initially located in the middle. We carried out a systematic study as a function of both the applied field and the excitation frequency. For each pair of values ($b,\omega $) we compute the time evolution of the system at $T=0$ during a time window $t_{sim}=500\,\tau$ using a time step $dt=5\times 10^{-3}\tau$ and with $\alpha=5\times 10^{-3}$ except at the right edge of the system, where a higher unrealistic value is considered ($\alpha=0.5$) to avoid SW reflection \cite{berkov_micromagnetic_2006}. For each simulation we obtain the total DW displacement and the power reflected by the DW. This latter one is estimated by integrating the time-averaged SW intensity ($m_z^2$) over the region to the left of the DW and subtracting from it the value computed in the same way but without the presence of the DW. The results of our study are presented in Fig.\ref{fig:DWdis-Ref}, where (a) the reflected power $P_r$ and (b) the DW displacement $\Delta x_{dw}$ are plotted as a function of both the applied field $b$ and excitation frequency $\omega$. The black lines in both figures correspond to the frequency of the uniform mode for each one of the two magnon branches ($\omega_+$ and $\omega_-$), where the values have been calculated using the analytical expression in \cite{rezende_introduction_2019} with $k=0$. We note that, unlike what happens in uniaxial AF, in our biaxial system ( eq.\ref{eq_hamiltonian}) the two modes are not degenerate at $b=0$ and there is a small gap between them. %On the other hand, it is worth mentioning that the sense of rotation imposed on the edge favours the excitation of the lower branch ($\omega_-$). If the opposite sense is imposed the upper branch ($\omega_+$) is favoured and propagation is only allowed for frequencies above the $k=0$ line, but even in this region the computed DW displacement is negligible in all cases.

A strong correlation between DW displacement and reflected power is apparent in Fig. \ref{fig:DWdis-Ref}, which indicates that magnon reflection at the DW is the dominant mechanism that explains the field-induced DW motion towards the colder regions. For a given frequency, as we increase the field (moving horizontally towards the right in Fig. \ref{fig:DWdis-Ref}) both $P_r$ and $\Delta x_{dw}$ increase monotonically. This can be explained by considering that, as already discussed, the DW net magnetic moment also increases with the field (Fig. 2(b)), which results into a higher probability for the magnons to be reflected, consequently transferring linear momentum to the DW an pushing it to the right \cite{shen_driving_2020}.

On the other hand, a strong frequency dependence in both $P_{r}$ and $\Delta x_{dw}$ is observed. For a given applied field the highest reflected power is obtained in a region slightly above the propagation threshold of the lower branch ($\omega_{-}$). This dependence is very closely related to the frequency dependent magnon propagation length in our system. Therefore, we assume that the reflected power in our case is mostly given by the SW intensity that reaches the DW as it propagates from the left edge of our system. The DW displacement shows a similar frequency dependence (Fig. \ref{fig:DWdis-Ref}(b)), although the maximum is slightly shifted towards higher frequencies. This shift is due to the fact that the magnon linear momentum is proportional to the wave vector $k$ \cite{shen_driving_2020} and therefore, the higher the frequency (wave vector), the higher the momentum transferred to the DW when the magnon is reflected. 

In view on this analysis with monochromatic SWs, we believe that for the thermally driven DW motion described in the first part of the section, it is also the low frequency magnons that contribute more significantly to the force driving the DW towards the cold end, not so much because their propagation length is larger but because their energy is smaller and, therefore, this region will be more populated when excited thermally. 

\section{\label{sec:conclusions}Conclusions}
We have shown that in AF, thermally driven DW motion towards the hotter regions can be reversed by applying a magnetic field along the easy axis. The domain wall gradually shifts from moving towards the hot end to moving towards the cold one, changing sign roughly mid-way to the spin-flop transition. Furthermore, the DW velocity dependence on the applied field collapses to a single curve when normalized to the temperature gradient. A frequency analysis revealed that this effect is mainly due to, on one hand, the increase of the DW reflectivity, intimately related to the sizable field-induced net magnetic moment in the DW and, on the other, the lowering of the propagation threshold induced by the field, which allows for a high population of thermally excited low-frequency magnons that push the wall when they are reflected.

Although we chose a biaxial AF for our study, the biaxial nature is not relevant for the effect reported. In fact, we obtain a very similar behaviour if we consider a uniaxial AF. On the other hand, we chose an AF with high anisotropy to have a narrow DW and, therefore, make the computational study less time consuming, but a similar effect of the magnetic field is also found in low anisotropy materials. Therefore, we believe that the reported effect of the magnetic field on thermally driven DW motion is quite universal and robust and that it could be confirmed experimentally in most common AF insulators, such as hematite, NiO, Cr$_2$O$_3$, etc. In our opinion, this finding might allow for new functionalities in spintronic devices based on the manipulation of DWs. 

\section*{Acknowledgements}
We acknowledge MAT2017-87072-C4-1-P from the (Ministerio de Econom\'ia y Competitividad) Spanish government and project SA299P18 from
	Consejer\'ia de Educaci\'on Junta de Castilla y Le\'on and the Youth Employment Operational Program of Castilla y Leon.

\bibliographystyle{apsrev4-1}
\bibliography{pAFS-arXiv}

\end{document}